\documentclass[12pt]{iopart}
\usepackage{graphicx}
\usepackage{caption}
\usepackage{subfig}
\usepackage{epsfig}
\usepackage{iopams}
\newcommand{\p}{\partial}

\newcommand{\wh}[1]{\widehat{#1}}

\newcommand{\vek}[1]{\mathbf{#1}}
\newcommand{\abs}[1]{\left|#1\right|}
\newcommand{\spar}{\shortparallel} 
\newcommand{\nablan}{\nabla_{\perp}}
\newcommand{\nablap}{\nabla_{\spar}}

\begin{document}
\title[]{Isotope effect on blob-statistics in gyrofluid simulations of scrape-off layer turbulence}
\author{O~H~H~Meyer$^{1,2}$ and A~Kendl$^{1}$}
\address{$^1$Institut f\"ur Ionenphysik und Angewandte Physik, Universit\"at
  Innsbruck, 6020 Innsbruck, Austria} 
\address{$^2$Department of Physics and Technology, UiT - The Arctic University of Norway,
9037 Troms\o, Norway}
\ead{ole.meyer@uibk.ac.at}
\begin{abstract}
We investigate time-series obtained from gyrofluid simulations in coupled edge/scrape-off layer turbulence characteristic for fusion edge-region plasmas.
Blob birth near the separatrix produces intermittent signals whose statistics depend on the ion mass of the reactor fuel, pointing towards overall slower dynamics for heavier
isotopes. We find that a recently established shot-noise stochastic model for scrape-off layer fluctuations coincides reasonably well with the numerical simulations performed in 
this contribution.
\end{abstract}
	
\submitto{\NF}
\maketitle


\section{Introduction}
There is clear experimental evidence that scrape-off layer (SOL) signals from Langmuir probes in tokamak devices may be adequately modelled by assuming that filament transport follows
a shot-noise process \cite{Garcia12, Garcia15, Garcia16, Theodorsen16, GarciaKSTAR, Garcia17}. Characteristics may be used to extrapolate SOL-widhts and heat-loads on the confining vessel wall and are of subtle importance for 
the progess of the fusion program. While understanding of particle- and heat transport by filaments has made rapid progress \cite{Krash1, Krash, Manz, Manz13, Madsen, Matthias} (for an
excellent review see \cite{DIp_review}), 
there still lacks explanation on how the resulting confinement properties are influenced by the fuel mass - an important issue as fusion reactors such as ITER are intended to run on deuterium-tritium
plasma mixtures, compared to most laboratory experiments, employing protium or deuterium fuel.
Contrary to what is to be expected from gyro-Bohm classical and neoclassical transport estimates,
experiments show that heavier hydrogen isotopes produce enhanced plasma confinement in tokamaks \cite{Bessenrodt, Hawryluk, Liu, Xu}. Recently, improved confinement
for increased ion mass has also been observed in a reversed field-pinch configuration \cite{Lorenzini15}. It has been suggested that enhanced shear-flow activity for heavier hydrogen isotopes may suppress edge turbulence favorably. 
Gyrokinetic theory \cite{Hahm} and simulations show that the ion mass may result in enhanced residual zonal-flow levels \cite{Bustos15, Guo15, GarciaJ17}. Gyrofluid simulations of edge turbulence have found
improved confinement for heavier hydrogen mixtures with only mild dependence of zonal flow activity on ion mass \cite{paper1}. 
In turbulence simulations of edge conditions, self-consistent development of zonal structures together with consistent boundary conditions along the magnetic field lines is necessary
to capture the birth of excess density filaments out of the turbulence (close to the separatrix) and its influence on mediating the radial propagation of these filaments, which may
account for a significant part of particle losses observed in the edge region \cite{LaBombard, Boedo, Nespoli17, Nielsen17}. Seeded filament motion has been shown to depend crucially on the ion mass,
with heavy ions moving slower \cite{paper2}. It is well worth to study the dynamical interplay between edge/SOL turbulence and its coupling to filament propagation.
In this contribution we compare the shot-noise model \cite{Garcia12} to time-series obtained from coupled edge/SOL gyrofluid turbulence simulations carried deeply into the nonlinearly saturated state
and investigate how characteristic statistical blob-parameters are influenced by the ion mass. 


\section{Gyrofluid model and computation} 
\label{model_section}

The present simulations are based on the gyrofluid
electromagnetic model introduced by Scott \cite{scott05b}.
In the local delta-$f$ isothermal limit the model consists of
evolution equations for the gyrocenter densities $n_s$ and parallel velocities
$u_{s  \spar}$ of electrons and ions, where the index $s$ denotes the species with
$s \in (\mathrm{e}, i)$:

\begin{eqnarray} 
& \frac{\rmd_{s} n_s}{\rmd t} = & - \nablap u_{s \spar} + \mathcal{K} \left(
\phi_{s} + \tau_{s} n_{s} \right),  
\label{density} \\
\wh{\beta} \; \frac{\p A_{\spar}}{\p t} + \epsilon_{s} & \frac{\rmd_s u_{s \spar}}{\rmd
  t} = & - \nablap \left( \phi_{s} + \tau_{s} n_{s} \right) + 2 \epsilon_{s}
\tau_{s} \mathcal{K} \left( u_{s \spar} \right) - C J_{\spar}.
\label{parvel} 
\end{eqnarray}
The plasma beta parameter
\begin{equation*}
\wh{\beta} = \frac{4 \pi p_{\rme}}{B^{2}} \left( \frac{q R}{L_{\perp}} \right)^{2},
\end{equation*}
controls the shear-Alfv\'{e}n activity, and
\begin{equation*}
C = 0.51 \frac{\nu_{\rme} L_{\perp}}{c_{0}} \frac{m_\rme}{m_{i 0}} \left( \frac{q R}{L_{\perp}} \right)^{2}, 
\end{equation*}
mediates the collisional parallel electron response for $Z=1$ charged hydrogen
isotopes.
The gyrofluid moments are coupled by the polarisation equation
\begin{equation} 
\label{poleq}
\sum_{s} a_{s} \left[ \Gamma_{1} n_s + \frac{\Gamma_{0} - 1}{\tau_{s}} \phi \right] = 0,
\end{equation}
and Ampere's law
\begin{equation}
- \nablan^{2} A_{\spar} = J_{\spar} = \sum_{s} a_s u_{s \spar}.
\end{equation}
The gyroscreened electrostatic potential acting on the ions is given by
\begin{equation*}
\phi_{s} = \Gamma_{1} \left( \rho_{s}^{2} k_{\perp}^ {2}  \right) \wh{\phi}_{\vek{k}},
\end{equation*}
where $\wh{\phi}_{\vek{k}}$ are the Fourier coefficients of the electrostatic potential.
The gyroaverage operators $\Gamma_{0} (b) $ and $\Gamma_{1} (b) = \Gamma_{0}^{1/2} (b)$ correspond to
multiplication of Fourier coefficients by $I_{0}(b) e^{-b}$ and 
$I_{0}(b/2) e^{- b/2}$, respectively, where $I_{0}$ is the modified Bessel
function of zero'th order and we have introduced the shorthand notation $b = \rho_{s}^{2} k_{\perp}^ {2}$. 
We here use approximate Pad\'{e} forms with $\Gamma_{0} (b) \approx (1 +
b)^{-1}$ and $\Gamma_{1} (b) \approx (1 + b/2)^{-1}$ \cite{dorland93}. 

The perpendicular $\vek{E} \times \vek{B}$ advective and the parallel derivative operators
for species $s$ are given by 
\begin{equation*}
\frac{\rmd_{s} }{\rmd t} = \frac{\p }{\p t} + \delta^{-1} \left\{ \phi_{s}, ~ \right\},
\end{equation*}
\begin{equation*}
\nablap  = \frac{\p }{\p z} - \delta^{-1} \wh{\beta} \left\{ A_{\spar}, ~ \right\},
\end{equation*}
where we have introduced the Poisson bracket as
\begin{equation*}
\left\{ f, g \right\} = \left( \frac{\p f}{\p x} \frac{\p g}{\p y} - \frac{\p f}{\p y} \frac{\p g}{\p x}  \right).
\end{equation*}

In local three-dimensional flux tube co-ordinates $[x,y,z]$, $x$ is a (radial) flux-surface
label, $y$ is a (perpendicular) field line label and $z$ is the position along
the magnetic field line.
In circular toroidal geometry with major radius $R$, the curvature operator is given by
\begin{equation*}
\mathcal{K} = \omega_{B} \left( \sin z \; \frac{\p}{\p x} + \cos z \; \frac{\p}{\p y} \right),
\end{equation*}
where $\omega_{B} = 2 L_{\perp} / R$,
and the perpendicular Laplacian is given by
\begin{equation*}
\nablan^{2} = \left( \frac{\p^{2}}{\p x^{2}} + \frac{\p^{2}}{\p y^{2}} \right).
\end{equation*}
Flux surface shaping effects \cite{kendl03,kendl06} in more general tokamak or
stellarator geometry on SOL filaments \cite{riva17} are here neglected for simplicity.

Spatial scales in each drift plane are normalised by the drift scale $\rho_0 = \sqrt{T_\rme
  m_{i0}}/\rme B$, where $T_\rme$ is a reference electron temperature, $B$ is
the reference magnetic field strength and $m_{i0}$ is a reference ion mass,
for which we use the mass of deuterium $m_{i0} = m_\mathrm{D}$. The parallel coordinate is normalised by the parallel connection length, $L_\spar = 2 \pi q R$, where $q$ is the safety factor at a
reference location inside the separatrix. The influence of the connection length on turbulence properties across the separatrix is studied in \cite{Ribeiro05}. 
The temporal scale is set by $c_0 / L_{\perp}$, where $c_0 = \sqrt{T_\rme/m_{i0}}$,
and $L_{\perp}$ is a perpendicular normalisation length (e.g. a generalized
profile gradient scale length), so that $\delta = \rho_0 / L_\perp$ is the drift scale. 
The temporal scale may be expressed alternatively $L_\perp / c_0 = L_\perp /
(\rho_0 \Omega_{0}) = (\delta \Omega_{0})^{-1}$, with the ion-cyclotron
frequency $\Omega_{0} = c_0 / \rho_0$.  In the following we employ $\delta = 0.015$ such that one normalised time
unit corresponds to $(\delta \Omega_0)^{-1} \sim 10^{-4}~ \mathrm{s}$. Fluctuation amplitudes are normalised according to 
$n_s / n_0 \rightarrow n_s$, $e \phi / T_\rme \rightarrow \phi$, $A_\spar L_\perp / \beta_\rme B \rho_0 q R \rightarrow A_\spar$,
$J_\spar L_\perp / e n_0 c_0 q R \rightarrow J_\spar$, $u_{s \spar} L_\perp / c_0 q R \rightarrow u_{s \spar}$ with the electron beta $\beta_\rme = 4 \pi p_0 / B^2$ in 
terms of the background electron pressure $p_0 = n_0 T_\rme$. Note that this normalisation produces the factor $\delta^{-1}$ with the Poisson brackets.

The main species dependent parameters are
\begin{eqnarray*}
a_{s} = \frac{Z_s n_{s0}}{n_{\rme 0}} , \quad \tau_{s} = \frac{T_{s}}{Z_s T_{\rme}}, \quad
\mu_{s} = \frac{m_{s}}{Z_s m_{i0}}, \\
\rho_{s}^{2} = \mu_{s}\tau_{s} \rho_{0}^{2}, \quad \epsilon_{s} = \mu_{s} \left( \frac{q R}{L_{\perp}} \right)^{2},
\end{eqnarray*}
setting the relative concentrations, temperatures, mass ratios and FLR scales
of the respective species. $Z_s$ is the charge state of the species $s$ with
mass $m_s$ and temperature $T_s$. Note that the index "$s$" denotes both electrons and ions, while the index "$i$" represents ion species such as
protium, deuterium, tritium or helium.

We employ $\wh{\beta} = 1$, $\omega_B = 0.04$, $\delta = 0.015$, $( q R / L_\perp)^{2} = 27,000$ and magnetic shear $\wh{s} = 1$, corresponding 
approximately to $L_\perp = 4.25 \mathrm{cm}$, $R = 165 \mathrm{cm}$, $B = 2 \mathrm{T}$, $T_\rme = 70 \mathrm{eV}$, $q = 3$ and $L_\spar = 31 \mathrm{m}$, typical for ASDEX Upgrade conditions 
close to the separatrix. Similar parameters for this numerical setup have been employed in \cite{kendl14}. The collisionality parameter is set to $C = 50$, since this is known to
increase radial blob velocity \cite{paper2}. Lower collisionality requires longer simulation times since fewer (and slower) blobs are produced. A similar trend with 
respect to collisionality has been found in \cite{Nespoli17, Nielsen17}. The ion temperature is fixed at $\tau_\mathrm{i} = 1$.

\subsection*{Parallel boundary conditions}

We distinguish between two settings for parallel boundary conditions in 
3-d simulations. In the case of edge simulations a toroidal
closed-flux-surface (CFS) geometry is considered, and quasi-periodic globally
consistent flux-tube boundary conditions in the parallel direction
\cite{scott98} are applied on both state-variables $n_\rme, \phi$ and flux
variables $v_{\rme \spar}, u_{s \spar}$.   

In the SOL domain, the state variables assume zero-gradient Neumann (sheath)
boundary conditions at the limiter location and the flux variables are given as 
\begin{eqnarray}
u_{s \spar}|_{\pm \pi} &= p_\rme|_{\pm \pi} = \pm \Gamma_d n_\rme|_{\pm \pi},\\
v_{\rme \spar} &= u_{s \spar}|_{\pm \pi} - J_\spar |_{\pm \pi}  = \pm \Gamma_d
[(\Lambda + 1) n_\rme|_{\pm \pi} - \phi|_{\pm \pi}], 
\end{eqnarray}
at the parallel boundaries $z = \pm \pi$ respectively \cite{Ribeiro05}. 
Note that in order to retain the Debye sheath mode in this isothermal model, the Debye current  
$J_\spar |_{\pm \pi} = \pm \Gamma_d (\phi - \Lambda T_\rme)$ is expressed as
$J_\spar |_{\pm \pi} = \pm \Gamma_d (\phi - \Lambda n_\rme)$ and the electron
pressure $p_\rme = n_\rme T_\rme$ is replaced by $p_\rme = n_\rme$ \cite{Ribeiro05}. 
This edge/SOL set-up and its effects on drift wave turbulence has been
presented in detail by Ribeiro \etal in Refs.~\cite{Ribeiro05,Ribeiro08}.

The sheath coupling constant is $\Gamma_d = \sqrt{(1 + \tau_i) / (\mu_i \wh{\epsilon})}$. 
The floating potential is given by $\Lambda = \Lambda_0 + \Lambda_i$, where
$\Lambda_0 = \log \sqrt{m_{i0} / (2 \pi m_\rme)}$ and $\Lambda_i = \log
\sqrt{\mu_i / (1 + \tau_i)}$. 
Here terms with the index $i$ apply only to the ion species. 
The expressions presented here are obtained by considering the finite ion
temperature acoustic sound speed, $c_i = \sqrt{(Z_i T_i + T_\rme) / m_i}$, 
instead of $c_0$ in Ref.~\cite{Ribeiro05}. This results in the additional
$\Lambda_i$, and the normalisation scheme yields the extra 
$\sqrt{(1 + \tau_i) / \mu_i}$ in $\Gamma_d$.

\subsection*{Numerical implementation}

Our code TOEFL \cite{kendl14} is based on the delta-$f$ isothermal electromagnetic
gyrofluid model \cite{scott05b} and uses globally consistent flux-tube
geometry \cite{scott98} with a shifted metric treatment of the coordinates
\cite{scott01} to avoid artefacts by grid deformation. In the SOL region a
sheath boundary condition model is applied \cite{Ribeiro05,Ribeiro08}. 
The electrostatic potential is obtained from the polarisation equation by an
FFT Poisson solver with zero-Dirichlet boundary conditions in the (radial)
$x$-direction. Gyrofluid densities are adapted at the $x$-boundaries to ensure
zero vorticity radial boundary conditions for finite ion temperature.  
An Arakawa-Karniadakis scheme is employed for advancing the moment
equations \cite{arakawa66,karniadakis91,Naulin}. 

\subsection{Shot-noise model}

The statistical model is outlined in detail in \cite{Garcia12,Garcia16,Theodorsen16}. The underlying assumption is that filament-propagation in the SOL is comprised of uncorrelated pulses of shape
$\psi(t)$ at time $t$ such that the particle density fluctuations ($\Phi (t)$) recorded at a single point is given by a superposition of arriving blobs
\begin{equation}
\Phi (t) = \sum_{k = 1}^{K (T)} A_k \psi (t - t_k),
\end{equation}
where a blob ($k$) of amplitude $A_k$ arrives at time $t_k$. During the total time of measurement, $T$, there arrive precisely $K$ pulses. The duration of each pulse
is given by
\begin{equation}
\tau_\rmd = \int_{-\infty}^\infty \rmd t \abs{\psi(t)}.
\end{equation}
Further assuming that pulses arrive according to a Poisson distribution, it follows that the waiting times between subsequent blob events is given by the exponential distribution.
The average waiting time is denoted by $\tau_\mathrm{w}$ and can be estimated by $\tau_\mathrm{w} = K / T$ or by fitting an exponential function to recorded values of 
waiting times. The stochastic process at hand then features the intermittency parameter $\gamma = \tau_\rmd / \tau_\mathrm{w}$, quantifying the degree of pulse overlap (strong overlap
for $\gamma \gg 1$ and vanishing overlap for $\gamma \ll 1$). As $\gamma \rightarrow \infty$, the probability density function for $\Phi$ approaches the normal distribution.
Blob shapes may be modelled according to
\begin{eqnarray} \label{pulse}
\psi (t) = \begin{cases}{
\exp(t / \tau_\mathrm{r}) ~~~~\mathrm{for}~ t < 0, \\
\exp(- t / \tau_\mathrm{f}) ~~\mathrm{for}~ t \geq 0, }\end{cases}
\end{eqnarray}
where the pulse consists of a trailing wake with rise time $\tau_\mathrm{r}$ and a steep front with fall time $\tau_\mathrm{f}$ such that the whole pulse lasts 
$\tau_\rmd = \tau_\mathrm{r} + \tau_\mathrm{f}$.
For exponentially distributed filament amplitudes, it follows \cite{Garcia16} that the stationary probability density function for the particle density is the gamma distribution
\begin{equation} \label{pdf}
P (\wh{\Phi}) = \frac{\gamma^{\gamma/2}}{\Gamma (\gamma)} \left( \wh{\Phi} + \gamma^{1/2} \right)^{\gamma - 1} \exp( - \gamma^{1/2} \wh{\Phi} - \gamma),
\end{equation}
where the mean value $\langle \Phi \rangle$ and standard deviation $\Phi_\sigma$ define the normalized variable
\begin{equation}
\wh{\Phi} = \frac{\Phi - \langle \Phi \rangle}{\Phi_\sigma},
\end{equation}
and the shape parameter may be found from the skewness, $\gamma = 4 / S^2$, or the flatness, $\gamma = 6 / (F - 3)$, of the raw signal. The stochastic model consequently
implies a parabolic relation between skewness and flatness, $F = 3 S^2 /2$, typically observed in experiments \cite{Theodorsen16, GarciaKSTAR}. Note that for a normal distribution $S = 0$ and $F = 3$.
The stochastic model also predicts an autocorrelation function for the case of two-sided exponential pulses such as in equation (\ref{pulse}) with $\tau_r = \lambda \tau_d$
and $\tau_f = (1 - \lambda) \tau_d$ \cite{Garcia17}:
\begin{equation} \label{acf}
\mathrm{R} (\tau, \lambda) = \frac{1}{1 - 2 \lambda} \left\{ (1 - \lambda) \exp \left[- \frac{\abs{\tau}}{(1-\lambda) \tau_d} \right] - \lambda \exp \left[- \frac{\abs{\tau}}{\lambda \tau_d}\right] \right\}.
\end{equation}
\section{Numerical simulations}
\begin{figure} 
    \includegraphics[width=10cm]{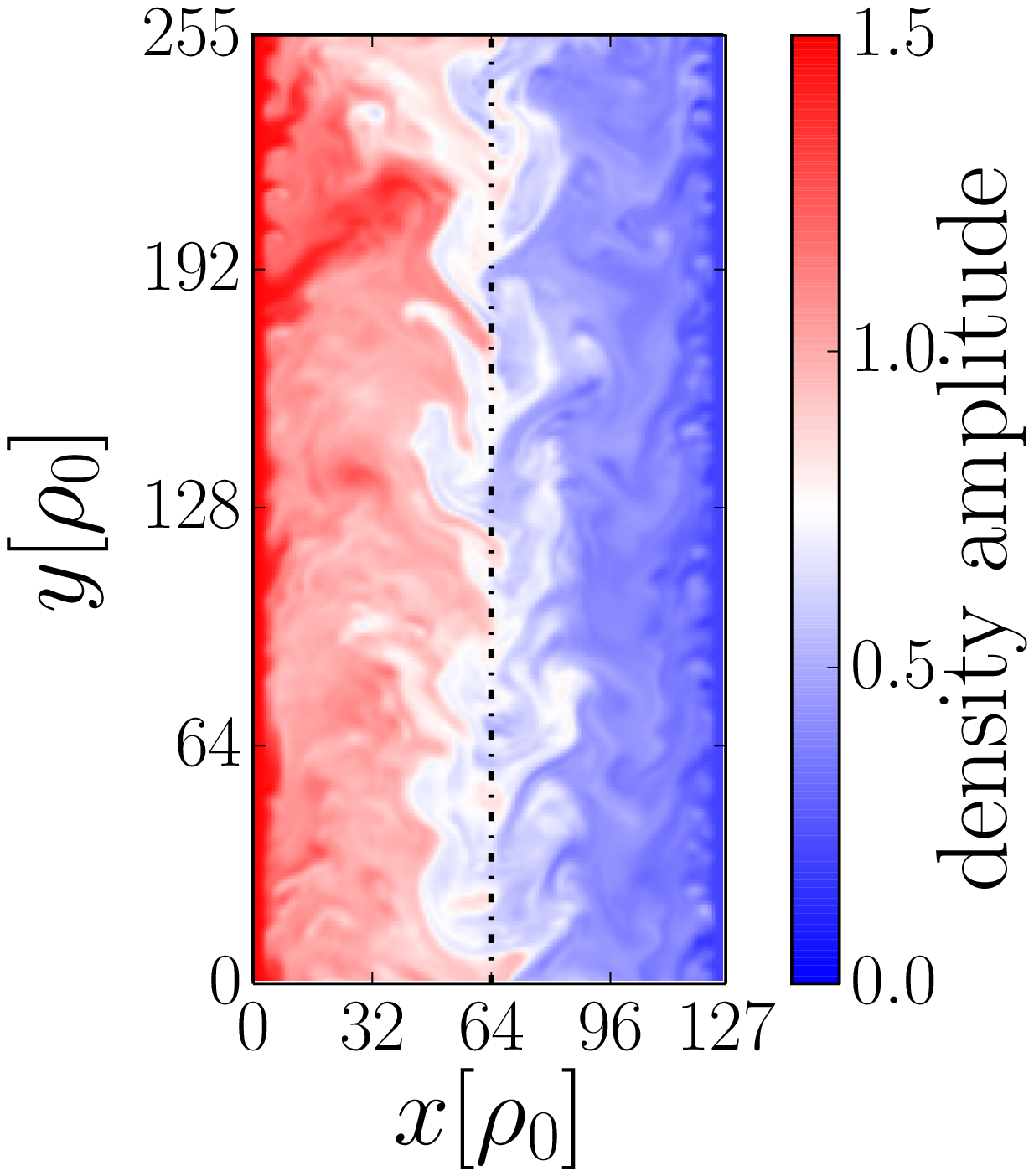} 
    \includegraphics[width=10cm]{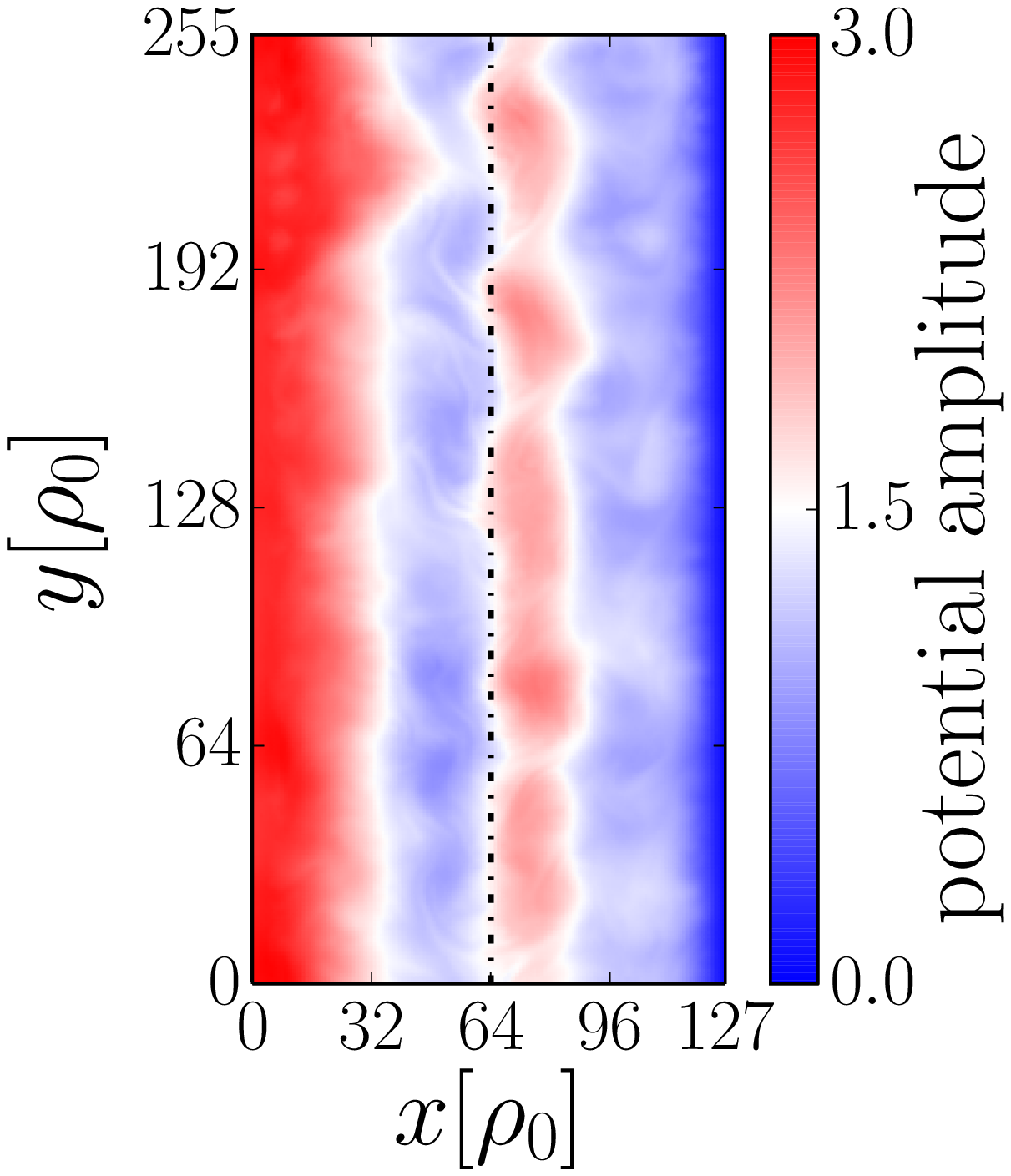} 
\caption{Snapshot of density (left) and potential (right) contours at the outboard midplane for deuterium ions.  }   
\label{contours}
\end{figure}

We chose to simulate a domain of dimensions $L_x \times L_y \times L_z = \left[128 \times 256\right] \rho_0  \times 2 \pi$ 
with resolution $128 \times 256 \times 8$. This corresponds to a box centered at the last closed magnetic flux-surface with approximate radial width $5$ cm and
$10$ cm length in perpendicular direction. Runs are taken to $T = 20,000$ normalised time-units with saturation occuring around $t = 1,000$. Statistics are taken over the 
saturated state. Typical blob birth near the separatrix, located at $x = 64$, is depicted in figure \ref{contours}, showing snapshots of the density (left) and potential (right) field at time $t = 1000$ for deuterium ions (left).

\subsection{Statistical tools}
\begin{figure} 
    \includegraphics[width=9cm]{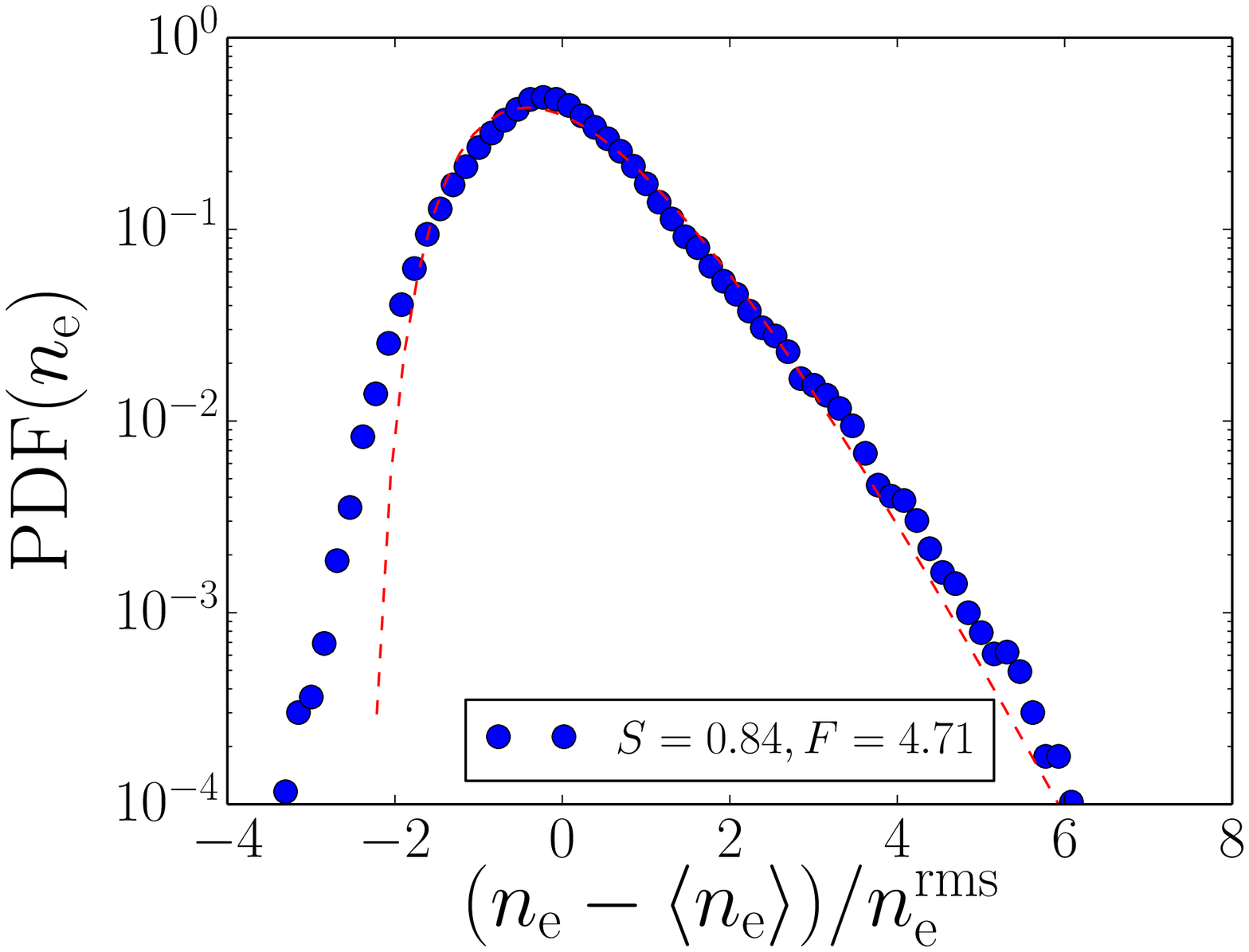} 
    \includegraphics[width=9cm]{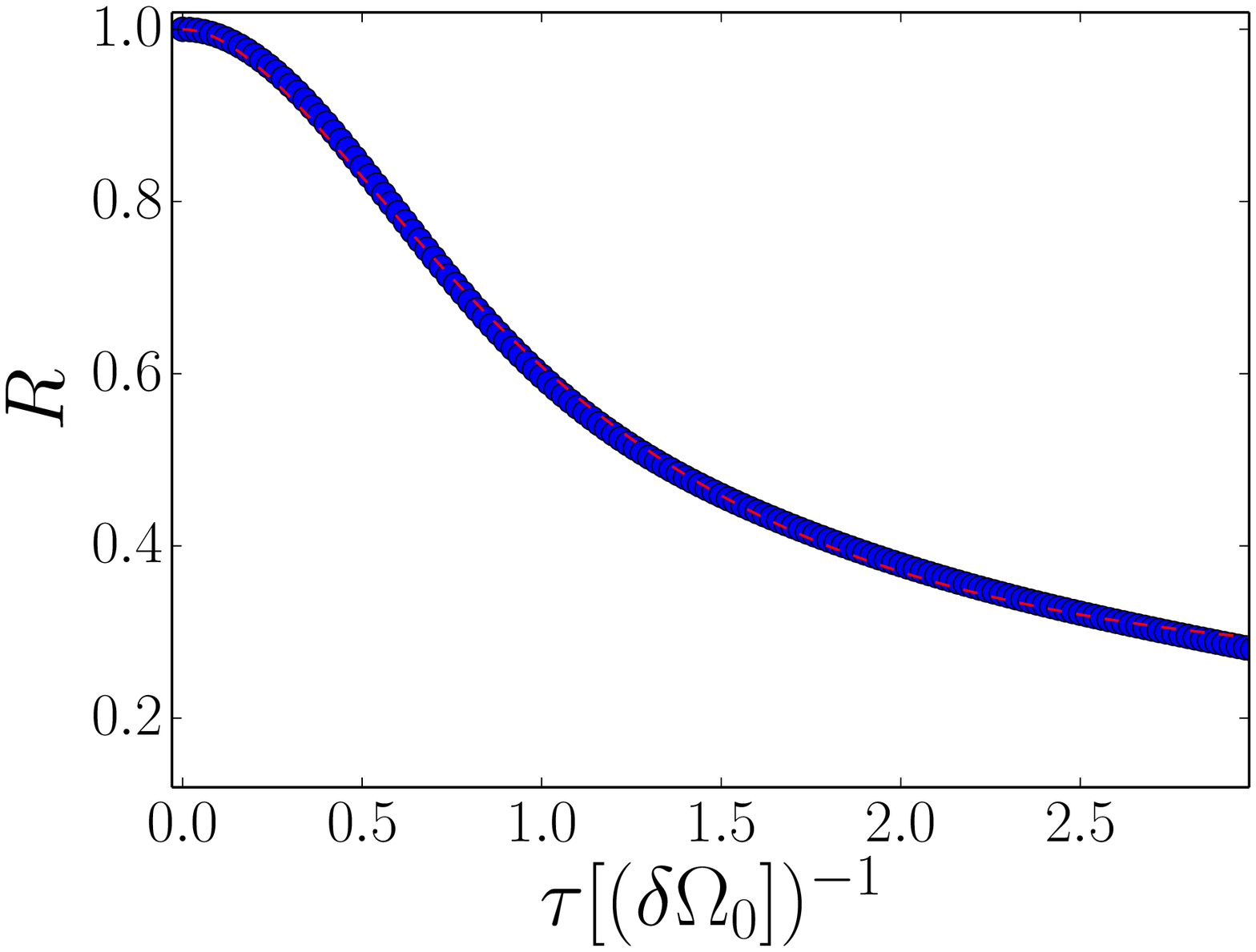} 
\caption{Probability density function (left) and autocorrelation function (right) for deuterium ions. The red dotted lines are fitted normalised gamma distribution, cf. eq. (\ref{pdf}), and theoretical
autocorrelation function, cf. eq.(\ref{acf}).}   
\label{pdf_acf}
\end{figure}

\begin{figure} 
    \centering
    \includegraphics[width=10cm]{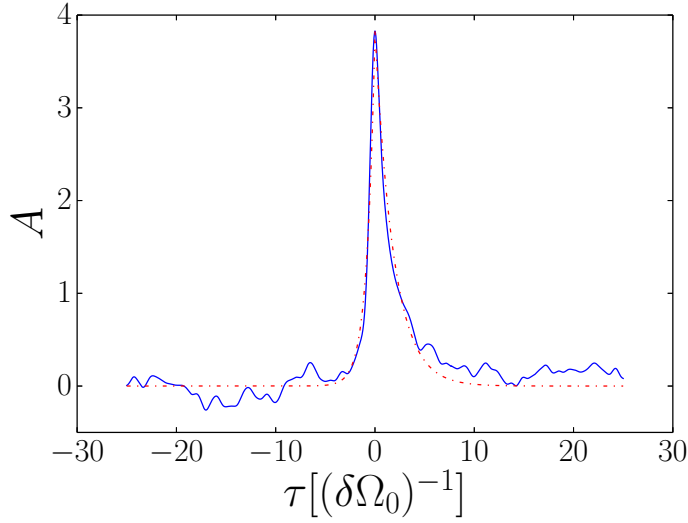} 
\caption{Conditionally averaged waveform for density fluctuations with deuterium ions.  }   
\label{cav}
\end{figure}

\begin{figure} 
    \includegraphics[width=9cm]{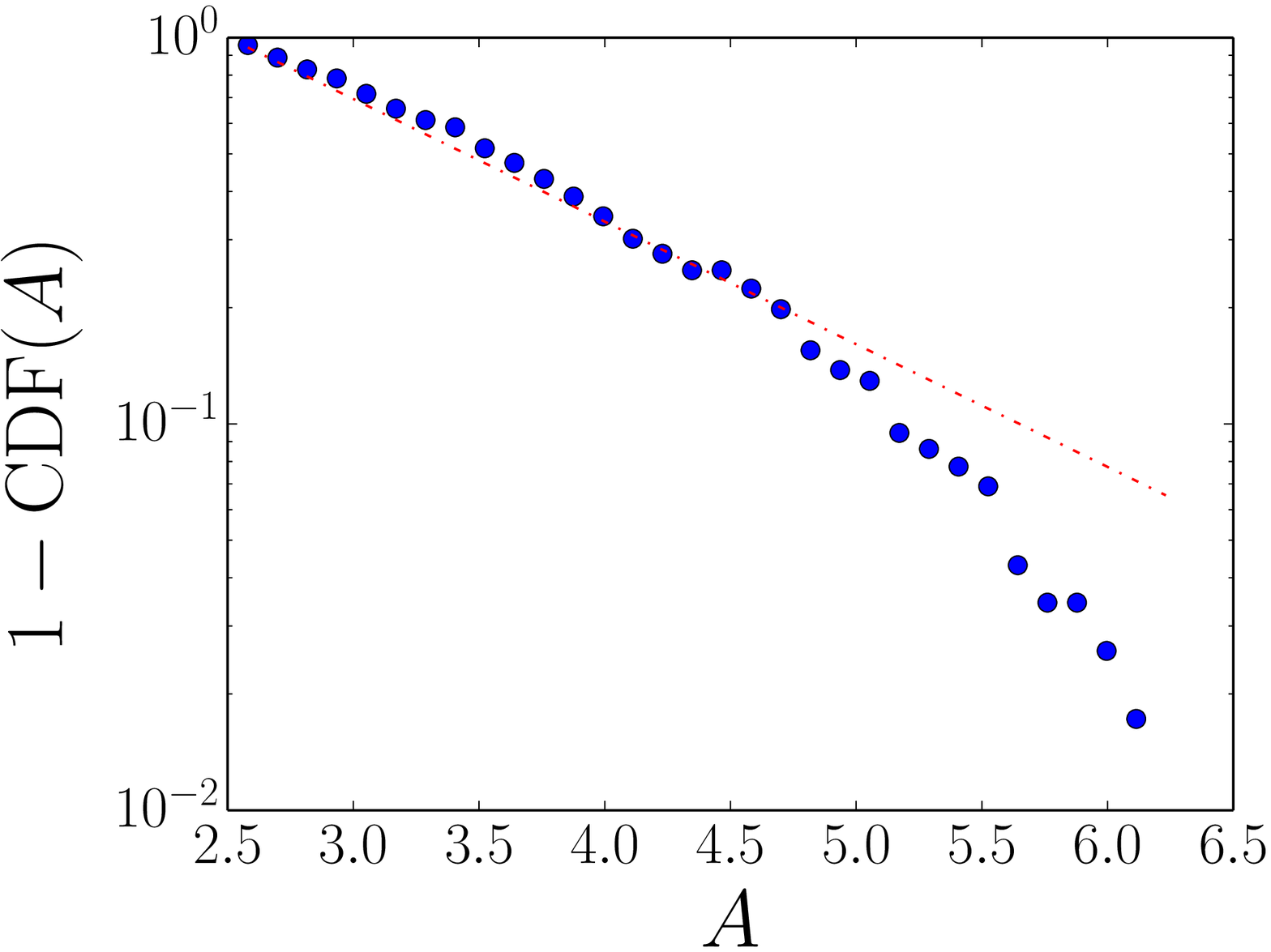} 
    \includegraphics[width=9cm]{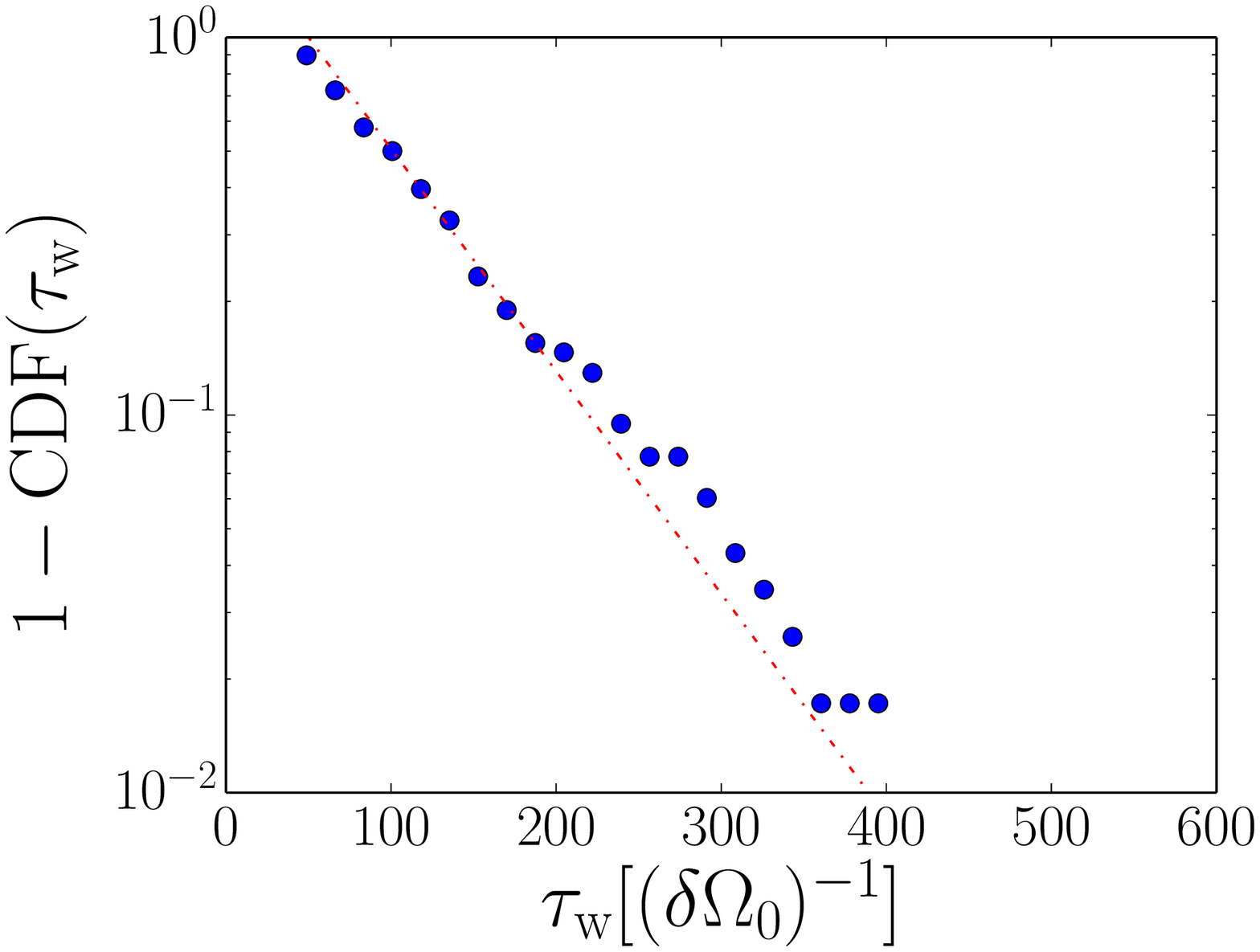} 
\caption{Complementary cumulative distribution functions for burst amplitudes (left) and waiting times (right) for deuterium ions. The red dotted line denotes fits from a truncated
exponential distribution.}   
\label{ccdfs}
\end{figure}

Figure \ref{pdf_acf} (left) shows the probability density function (PDF) for the electron particle density measured at the outboard midplane at $y = 128$ and $10$ grid points off the outermost
radial boundary, i.e. at $x = 117$. The fitted normalised gamma distribution suggested from the stochastic model provides a reasonable fit, at least for the tail of the PDF. Similar to the experimental findings in
\cite{GarciaKSTAR}, the autocorrelation function does not decrease to zero such that the actual fit employed in figure \ref{pdf_acf} (right) results from $R = c + (1 - c) \mathrm{R}$, where $c$ is the offset from zero and $\mathrm{R}$ is given
in equation (\ref{acf}). The fitted duration time is found to be $\tau_d = 1.1$.
Figure \ref{cav} shows the corresponding conditionally averaged wave-forms for filament 
bursts with peak ampltiude larger than $2.5$ times the averaged fluctuation level, where the density signal is normalised according to
\begin{equation}
\wh{n} = \frac{n_\rme - \langle n_\rme \rangle}{n^\mathrm{rms}_\rme},
\end{equation}
and the conditionally averaged amplitudes are recorded for distinct peak events such that $A = \wh{n} | \wh{n} > 2.5$. Fitting the pulse shapes in equation (\ref{pulse}) gives $\tau_r = 0.8$
and $\tau_f = 1.9$. For more details on conditional averages cf. \cite{Johnsen87, Huld91, Nielsen96, Oeynes95}. 
Average burst amplitudes and waiting times are calculated by fitting truncated exponential distributions to the complementary cumulative probability density functions in figure \ref{ccdfs}.
Note that for a stochastic variable, $X$, distributed according to an exponential distribution, the expectation value for events with amplitude larger than a threshold, $y$, is given
by 
\begin{equation}
\mathrm{E} (X | X > y) = y + \langle X \rangle.
\end{equation}
In figure \ref{ccdfs} the threshold is $y = 2.5$ for the blob amplitudes giving a mean value of $3.8$, in accordance with the maximum of the conditionally averaged signal in figure \ref{cav}.
The window length in figure \ref{cav} is $50$ time steps, such that the truncation threshold for the waiting time fit is $y = 50$, yielding an averaged waiting time of $138$. The deviation
from the fit function is likely due to few blob events at large amplitude or waiting time. Complementary simulations with smaller box size that were taken to $T = 40,000$ reveal that
more events produce a better fit.
\subsection{Ion mass scan}
\begin{figure} 
    \includegraphics[width=9cm]{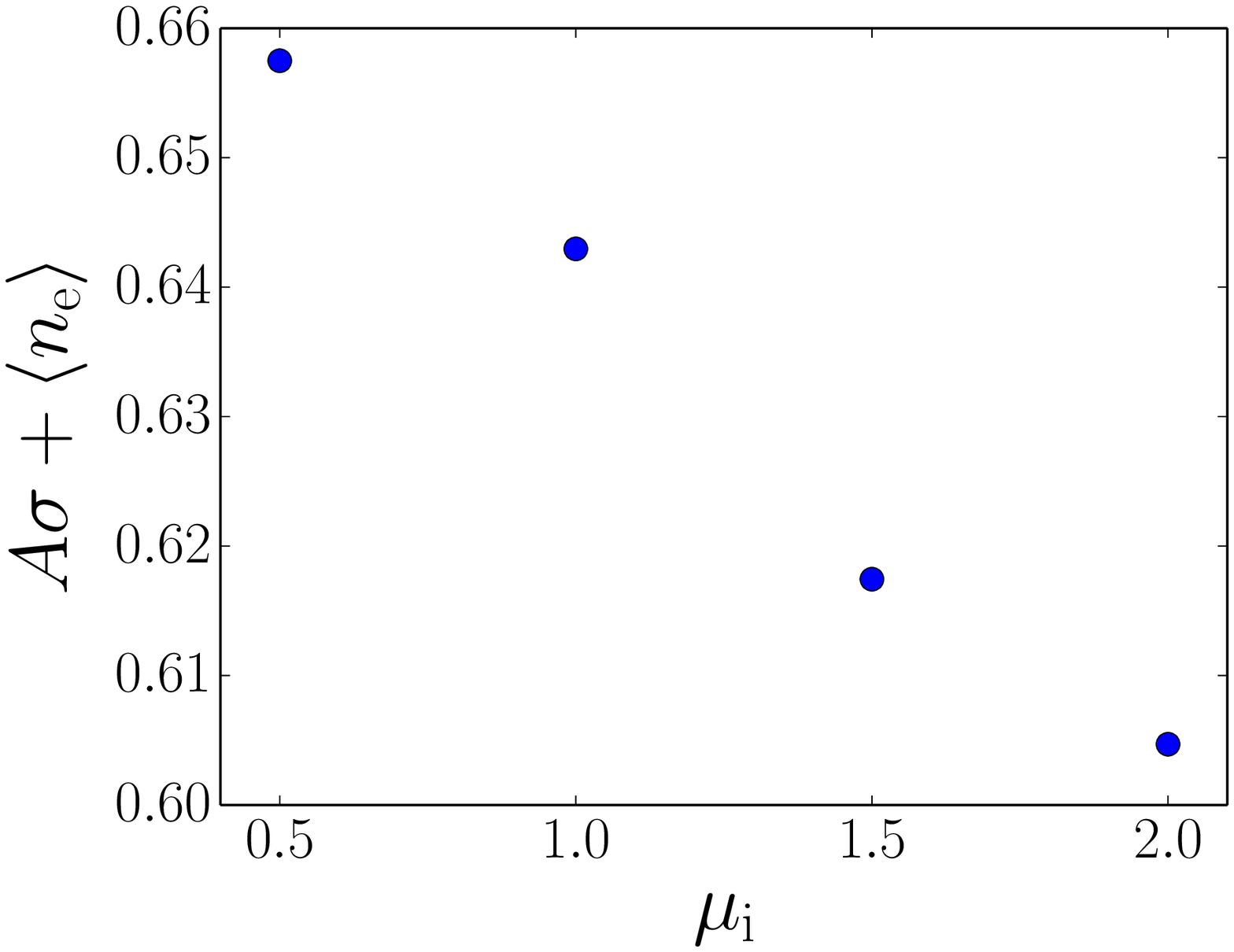} 
    \includegraphics[width=9cm]{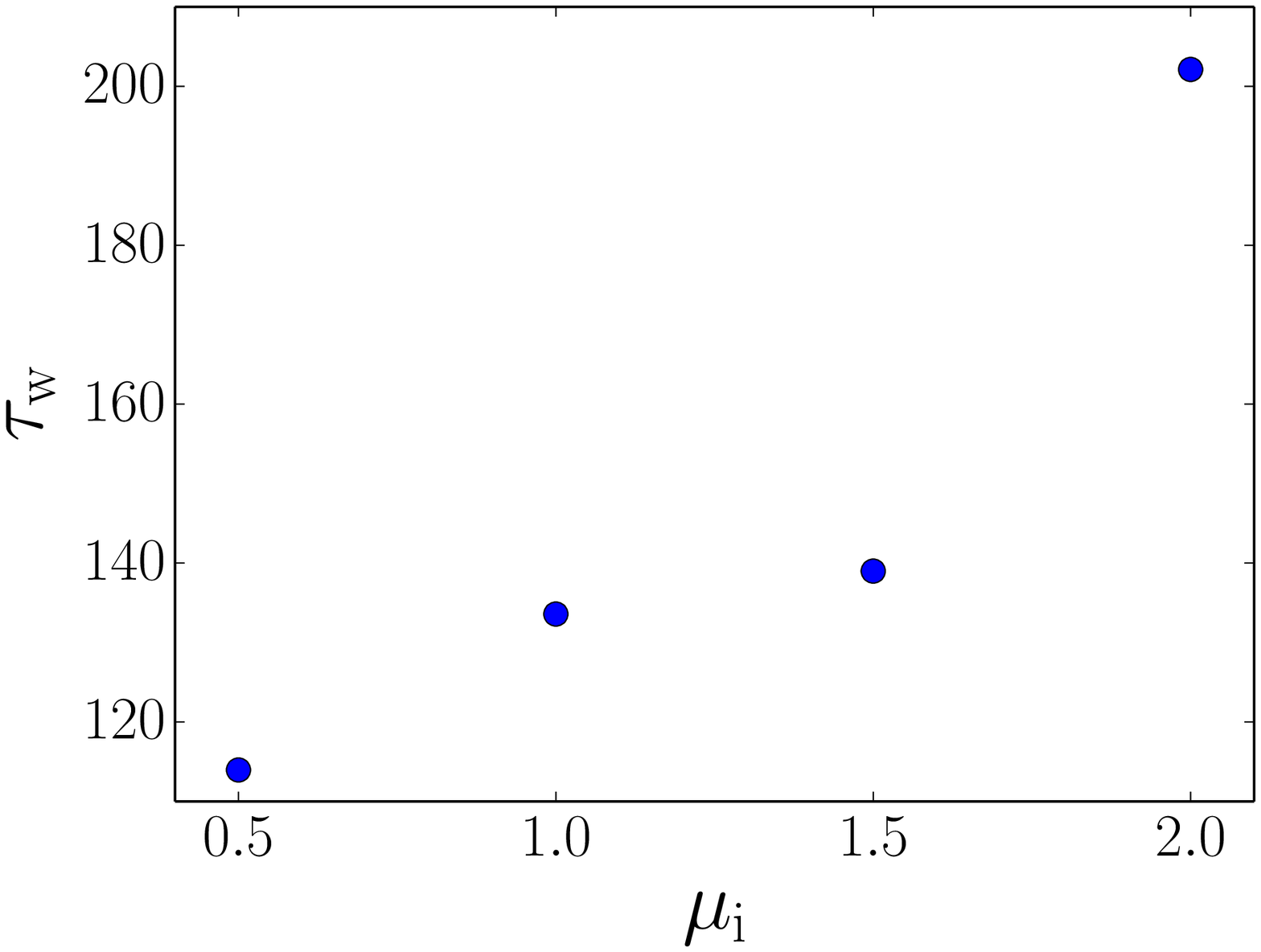} 
\caption{Average burst amplitude for the raw density signal (left) and waiting time (right) dependence on ion mass. }   
\label{amp_wait}
\end{figure}

\begin{figure} 
    \includegraphics[width=9cm]{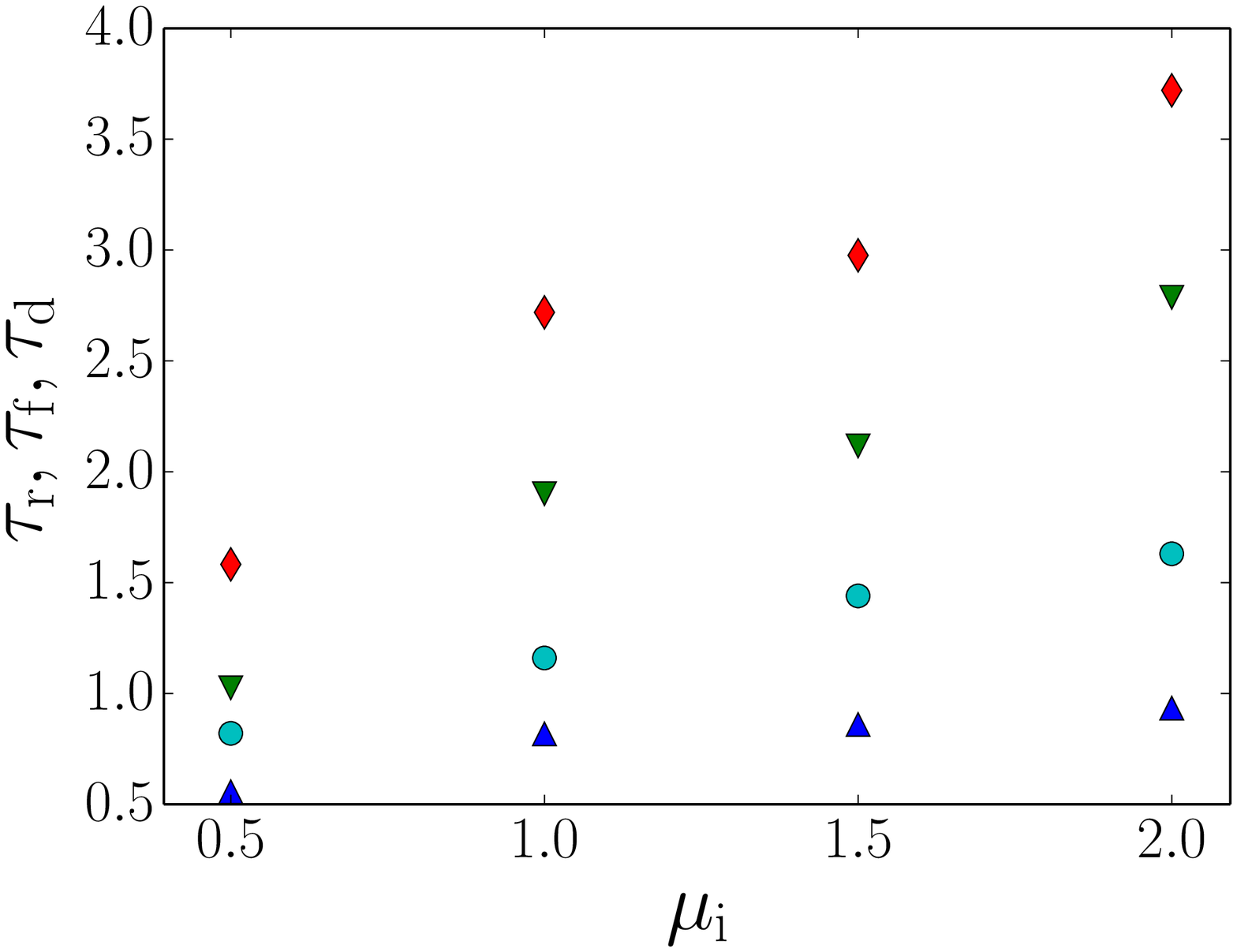} 
    \includegraphics[width=9cm]{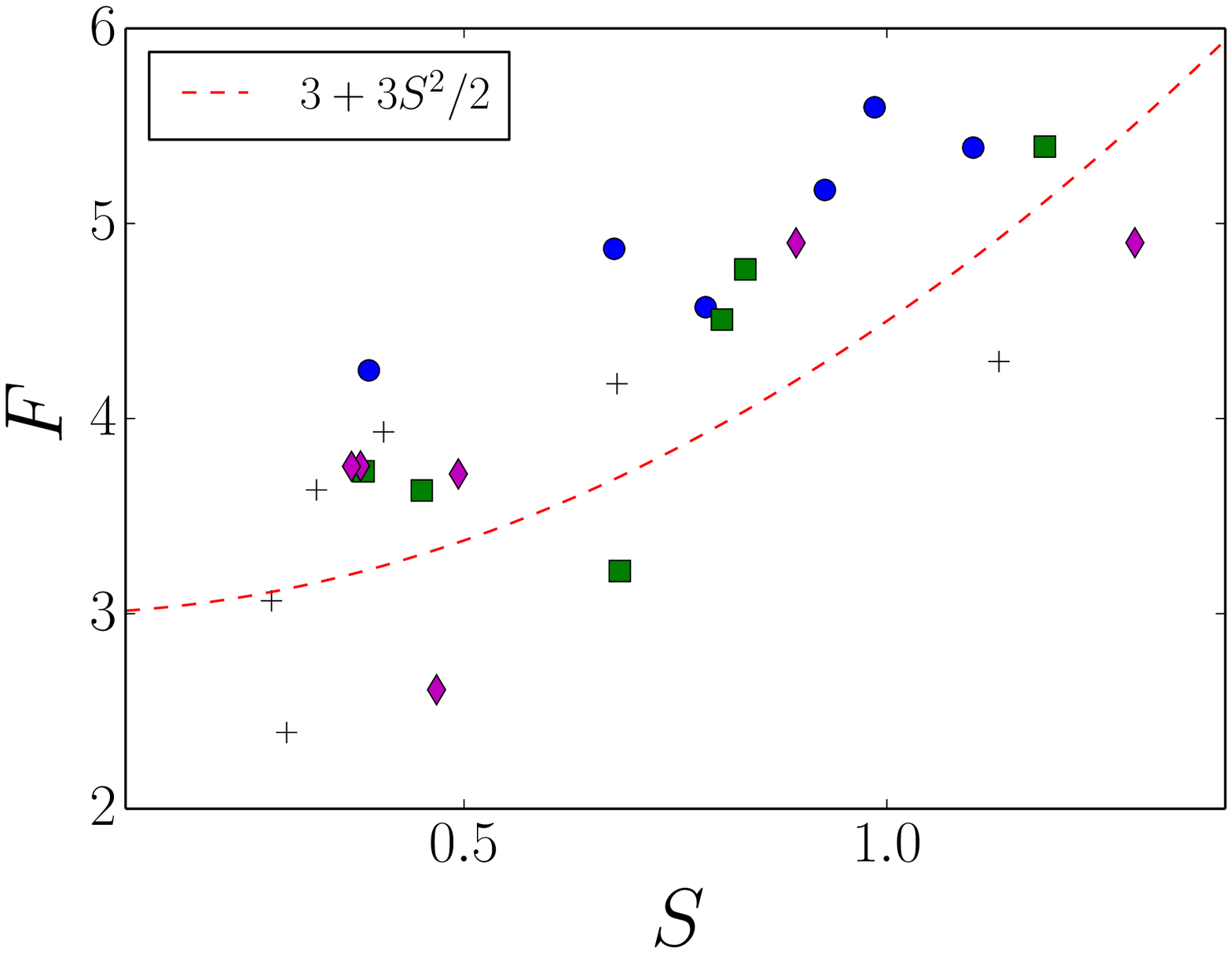} 
\caption{Left: Rise (blue upward triangles), fall (green downward triangles) and duration times computed from conditionally averaged waveforms (red diamonds) and autocorrelation functions (turquoise circles) dependent on ion mass. Right: Scatter
plot of flatness and skewness from varying radial positions in the SOL for varying ion mass (protium: blue circles, deuterium: green squares, tritium: magenta diamonds and 
singly charged helium: black crosses). The red dotted line marks the theoretical prediction from the stochastic model.}   
\label{tau_F_S}
\end{figure}

\begin{figure} 
    \includegraphics[width=9cm]{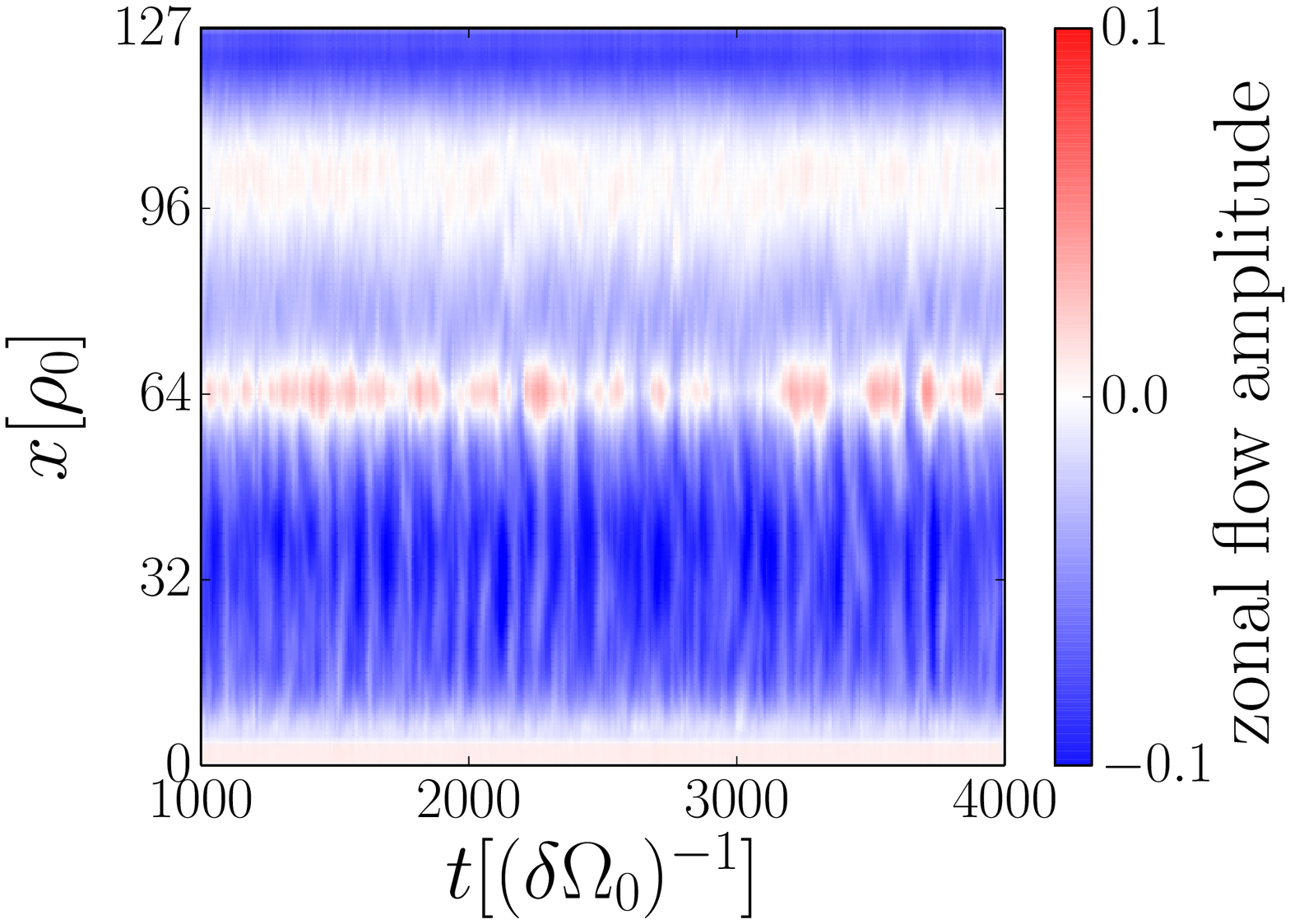} 
    \includegraphics[width=9cm]{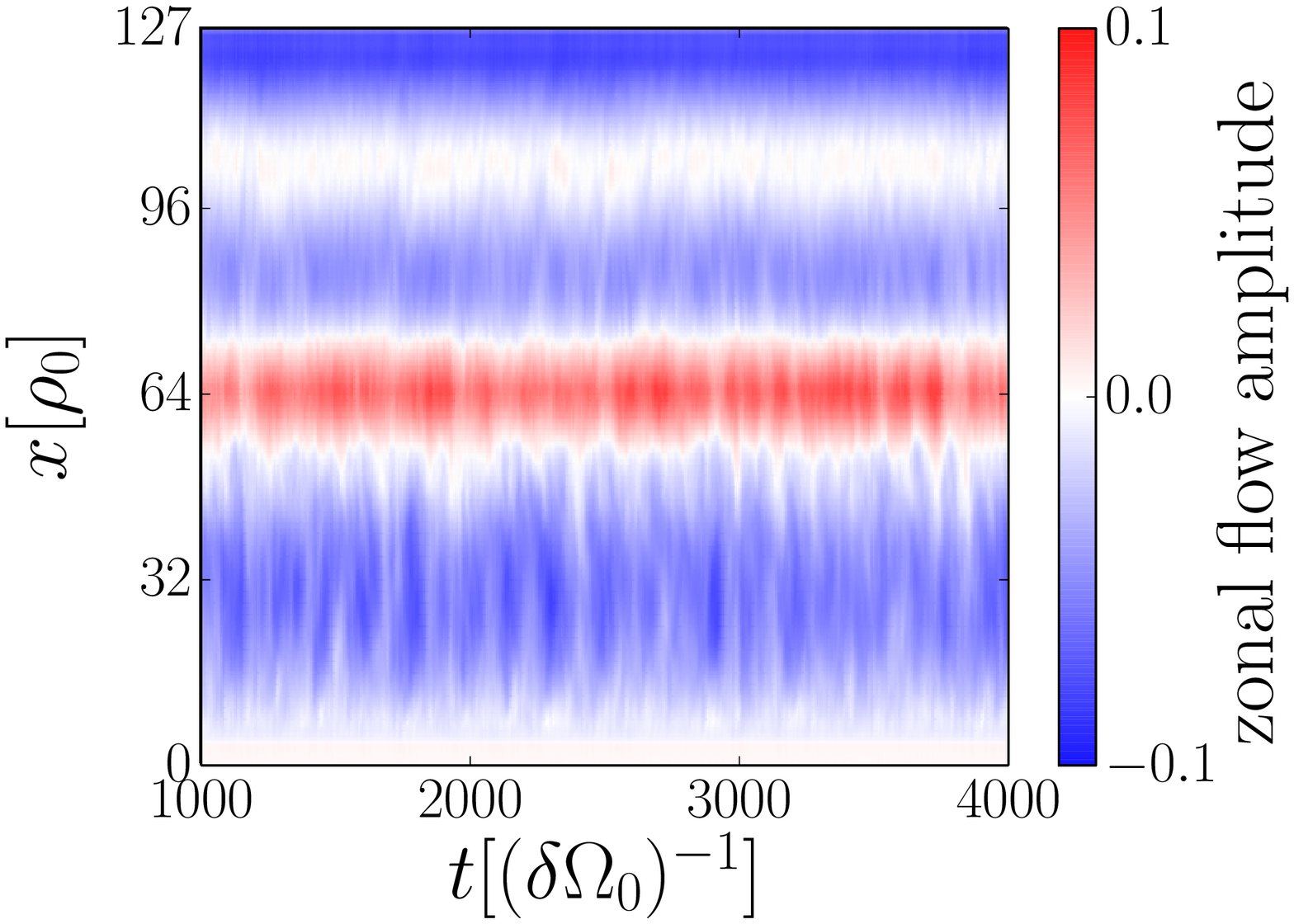} 
    \includegraphics[width=9cm]{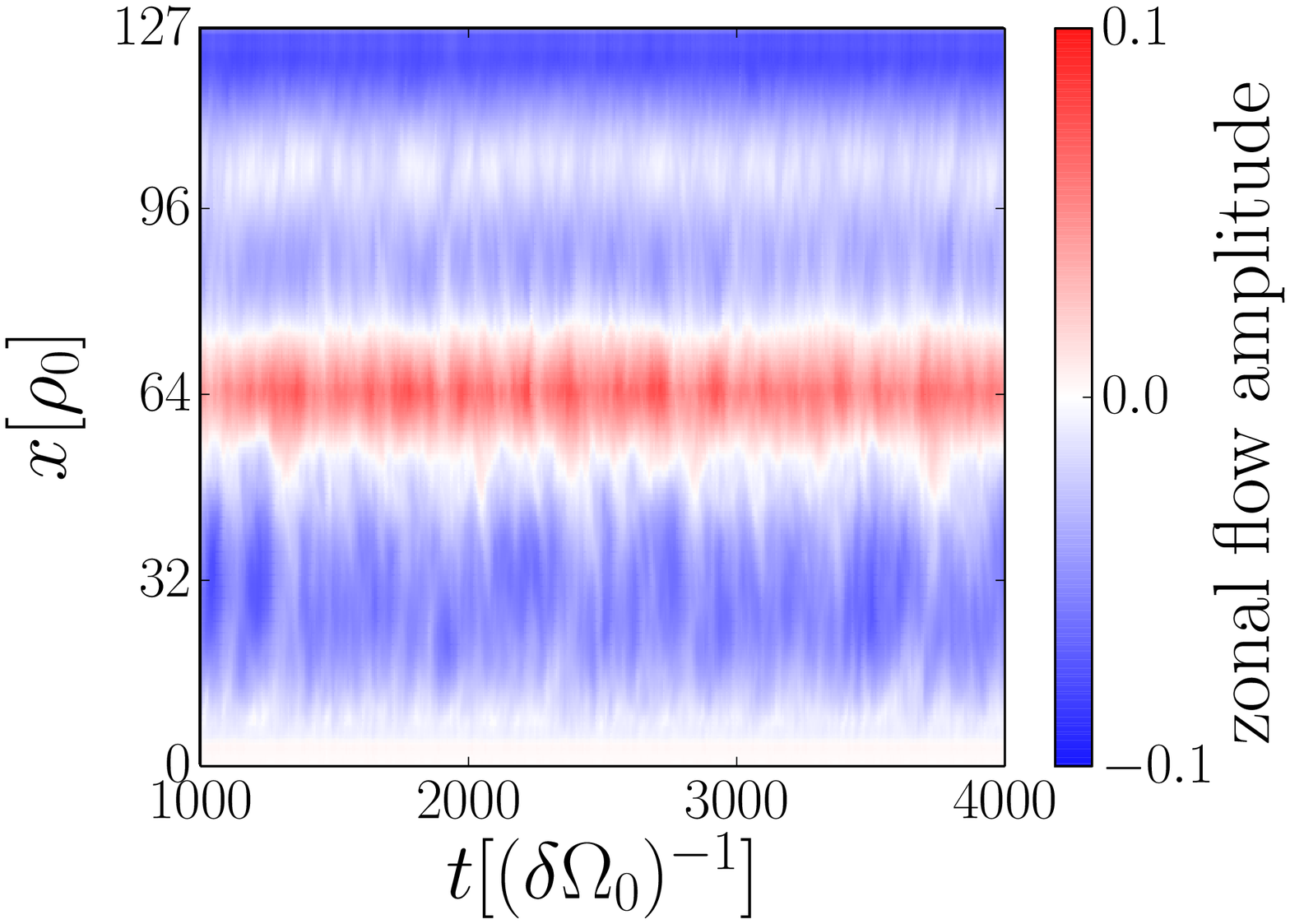} 
    \includegraphics[width=9cm]{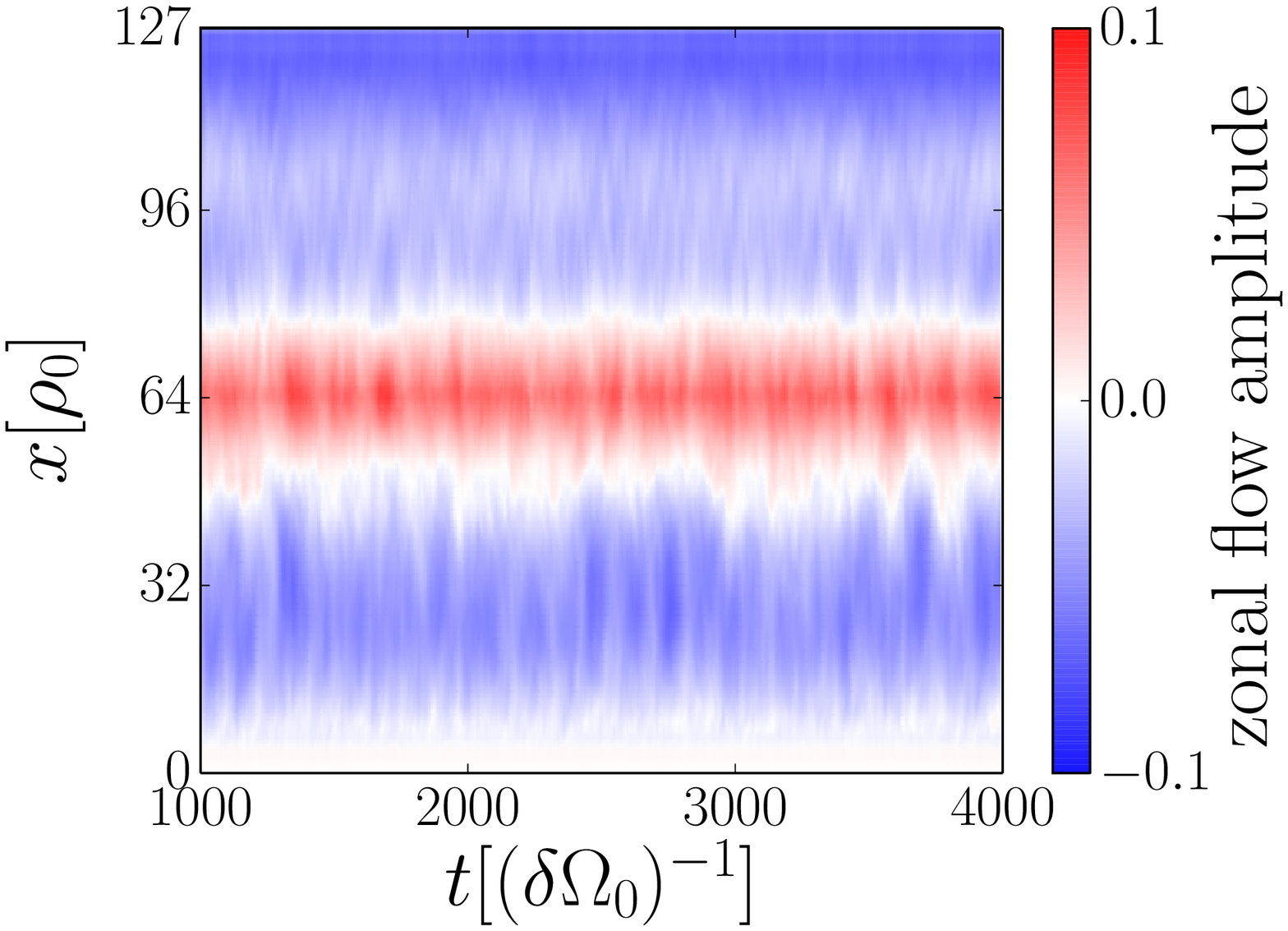} 
\caption{Flux-surface averaged flow velocity time evolution. Ions are: protium (top left), deuterium (top right), tritium (bottom left) and singly charged helium (bottom right).  }   
\label{zf_contours}
\end{figure}

In the same fashion we now employ the presented statistical techniques to signals obtained for setting the ion mass ratio, $\mu_\mathrm{i} = m_s / Z_s m_\mathrm{i0}$, to
protium ($\mu_\mathrm{i}=1/2$), deuterium ($\mu_\mathrm{i}=1$), tritium ($\mu_\mathrm{i}=3/2$) and singly charged helium ($\mu_\mathrm{i}=2$) values. Figure \ref{amp_wait} (left)
gives the renormalized density bursts obtained from computing the conditionally averaged waveform, fitting a truncated complementary cumulative exponential distribution to the recorded conditional amplitudes
and renormalizing by multiplying with the respective rms value and adding the mean value of the raw signal. Conditionally averaged peak amplitude show no trend with respect to ion mass, however,
the renormalized bursts do in terms of reduced bursts for higher ion mass. Similarly, we record the waiting times between conditional peaks and fit a truncated complementary cumulative exponential distribution, resulting
in figure \ref{amp_wait} (right), clearly showing that bursts are less frequent for higher ion mass.
In figure \ref{tau_F_S} (left) we present results from fitting the two-sided exponential pulse shape from equation (\ref{pulse}) to the conditionally averaged waveforms. We obtain 
estimates for the rise time, $\tau_r$, fall time, $\tau_f$ and total pulse duration, $\tau_r + \tau_f$. Furthermore, the fits to autocorrelation functions give a second estimate of the pulse duration.
Increasing ion mass produces pulses of longer rise, fall and duration according to both analyses. The right part of figure \ref{tau_F_S} shows for all runs at hand a scatter plot
of calculated skewness and flatness at $6$ distinct radial locations, separated by $10 \rho_0$, at $y = 64$ and the outboard midplane throughout the SOL region of the simulation domain. According to the stochastic model, 
there should be a parabolic relation between skewness and kurtosis. The figure \ref{tau_F_S} (right) shows that this indeed is a reasonable suggestion for the data presented, also indicating that this relation is universal in the sense that
it does significantly depend on the ion mass. Furthermore, the intermittency parameter, $\gamma = \tau_d / \tau_\mathrm{w} \in [0.01, 0.28]$, showing no clear trend with 
respect to isotopic composition. Signals are thus characterised by non-overlapping burst events.
The isotopic dependence on the zonal flow velocity is depicted in figure \ref{zf_contours}. Heavier isotopes produce zonal flows of slightly increased velocity and wavelength in 
the radial direction. Furthermore, it should be noted that the zonal flow for the protium simulation is not persistent in time. Computing the zonal-flow shear, 
$\partial_x^2 \langle \phi \rangle$, where $\langle \phi \rangle$ is the flux-surface averaged potential, we find no dependence on ion mass, correlating with the observed parameter
denpencies. It should be noted that the simulation presented here can be though of as typical for an L-mode scenary, hence the zonal flow is not dominating the turbulence.
\section{Conclusion}
We have shown that a recently established statistical model for filament propagation in the SOL of fusion plasmas may be used to describe time-series for gyrofluid turbulence
simulations that include self-consistent parallel boundary conditions to mimic the edge/SOL transition. The simulations show that burst amplitudes and waiting times follow an exponential distribution,
indicating that pulse arrival follows a Poisson process. Probability density functions are reasonably well approximated by gamma distributions and postulated autocorrelation functions
are a good description of the measured signals. A parabolic relation between skewness and flatness moments seems to be present. All of this is consistent with the assumptions behind the stochastic blob model.
It is shown that resulting electron density fluctuations close to the outermost
radial boundary at the outboard midplane feature an improved confinement state in terms of reduced amplitude bursts for heavier isotopes. Typical burst events are shown to be
of longer duration, with longer rise and fall times for increased ion mass. Regardless of ion mass the intermittency parameter is small, characteristic for non-overlapping
pulse arrival. The detection frequency of conditional blob events is higher for lighter isotopes.  
It is a well established fact that reduced blob detection frequency is associated with more pronounced shearing activity to break up radial streamers and decorrelate filaments \cite{DIp_review}.
According to \cite{Burrell} and references therein, shearing influences the perpendicular diffusion coefficients if $\omega_E > \gamma_L$, where $\omega_E \sim E_r \sim \partial_x^2 \langle \phi \rangle$ is the $\vek{E} \times \vek{B}$ shearing rate,
and $E_r$ the radial electric field. $\gamma_L$ is the maximum linear growth rate of the system. For either drift-wave or interchange turbulence, $\gamma \sim \mu_\mathrm{i}^\alpha$, with $\alpha = - 1$ 
for drift-wave instability and $\alpha = -3/4$ for interchange instability. The simulation at hand produces approximately constant values of maximum $\partial_x^2 \langle \phi \rangle$ with respect to ion mass,
hence, the ratio $\omega_E / \gamma_L \sim \mu_\mathrm{i}^\beta$ with $\beta > 0$, providing at least a qualitative argument that the shear flow dynamics may favorably influence the confinement improvement
in terms of reduced detection frequency for heavier isotopes.
Another explanation may be reduced
radial propagation velocities for heavier isotopes such as found in \cite{paper2}, together with overall slower dynamics (increased autocorrelation time), cf. \cite{paper1}, for edge turbulence, producing 
fewer blobs in a given time for increased ion mass. 
Blob amplitudes, skewness, flatness and intermittency seem to be universal in the sense that they do not depend on the ion mass.

Further studies, not making smallness assumpations on the relative amplitude of perturbations compared to the background should be executed, preferably through a full-f 3D gyrofluid
(or gyrokinetic) computational model in order to check the statistical dependency of blob properties \cite{Matthias, Nespoli17, Held16, Kendl15}.

\section*{Acknowledgments}
We acknowledge main support by the Austrian Science Fund (FWF) project Y398 and by the Austrian Academy of Sciences (OeAW) MG2016-6.
This work has been carried out within the framework of the EUROfusion
Consortium and has received funding from the Euratom research and training
programme 2014-2018 under grant agreement No 633053. The views and opinions
expressed herein do not necessarily reflect those of the European Commission.  
We gratefully acknowledge that A Theodorsen provided the script for calculating conditionally averaged signals.


\end{document}